\documentclass[12pt]{article}

\usepackage[margin=1in]{geometry}
\usepackage{graphicx}
\usepackage{amsmath,amssymb}
\usepackage[numbers,sort&compress]{natbib}
\usepackage[table]{xcolor}
\usepackage{booktabs}
\usepackage{enumitem}
\usepackage{tabularx}
\usepackage{adjustbox}
\usepackage{siunitx}
\usepackage{lmodern}

\graphicspath{{images/}}

\newcommand{\scen}[3]{#1 / \textbf{#2} / #3}
\definecolor{s1dom}{RGB}{225,245,254}
\definecolor{s2dom}{RGB}{255,243,224}

\title{The Hidden Water Geography of U.S. Hyperscale Data Centers in the AI Era}
\author{Gianluca Guidi$^{1,*}$ and Francesca Dominici$^{1,*}$}
\date{}

\begin{document}

\maketitle

\noindent$^{1}$Harvard University.\\
$^{*}$Correspondence to: ggianluca@hsph.harvard.edu; fdominic@hsph.harvard.edu.

\bigskip
\noindent\textbf{One-sentence summary:} Separating on-site cooling water from electricity-related water reveals that U.S. data centers face two geographically distinct water problems requiring different governance responses.

\begin{abstract}
Water use by data centers is routinely reported as a single footprint, but water is consumed through two physically distinct pathways: at the site for cooling and in the power system that generates electricity. We mapped both pathways for 472 U.S. hyperscale facilities by linking facility locations to electricity regions, hydrologic basins, and water-stress data. Under baseline assumptions, operational water consumption totals approximately 300~GL~yr$^{-1}$ (range 205--451 across scenarios), with electricity-related water contributing three-quarters of the total. The two pathways produce different hotspot geographies: direct cooling burdens concentrate in stressed western and south-central basins, whereas electricity-related burdens concentrate in a few eastern grid regions with fossil-heavy supply. Just 3 of 24 hosting balancing authorities account for 59\% of electricity-related water. Separating pathways identifies which decisions matter where: cooling design and water sourcing locally, electricity planning and procurement regionally.
\end{abstract}

\section*{Main Text}

Hyperscale data centers are expanding rapidly as cloud services and artificial intelligence workloads grow. Their electricity demand has received substantial attention from utilities and regulators, but their water use remains harder to measure in a way that guides decisions. The difficulty is not only estimating how much water is consumed. It is also identifying \emph{where} that water is consumed and \emph{which institutions} can reduce it.

Operational water consumption arises through two pathways. Water is consumed at the facility for thermal management, mainly through evaporative cooling; we refer to this as Scope~1, or direct cooling water \citep{mytton2021datacentre,iso30134-9,epaWatersenseAtWork}. Water is also consumed in the electricity system that serves the facility, including thermoelectric cooling at fossil-fuel and nuclear plants and, depending on attribution choice, reservoir evaporation associated with hydropower; we refer to this as Scope~2, or electricity-related water \citep{macknick2012water,meldrum2013lifecycle}. Existing studies have advanced understanding of each pathway individually \citep{mytton2021datacentre,li2023thirsty,li2025thirsty,ristic2015water,shehabi2016united}, but these pathways have not been mapped together in a facility-resolved national analysis. Collapsing them into one footprint hides which problem is local and which is regional.

Place matters because the same volume of water consumption can imply different levels of risk depending on local hydrology and baseline water stress \citep{wriAqueduct,kuzma2023aqueduct40}. Cooling-water intensity also varies with climate, cooling architecture, and operating efficiency \citep{lei2022climate,shehabi2016united}. National or state totals alone are therefore insufficient for planning. A site may appear water-light locally but be connected to a water-intensive grid, or it may impose direct cooling demand in a stressed basin even when the electricity supply is less water-intensive.

\begin{figure}[t]
\centering
\includegraphics[width=0.95\textwidth]{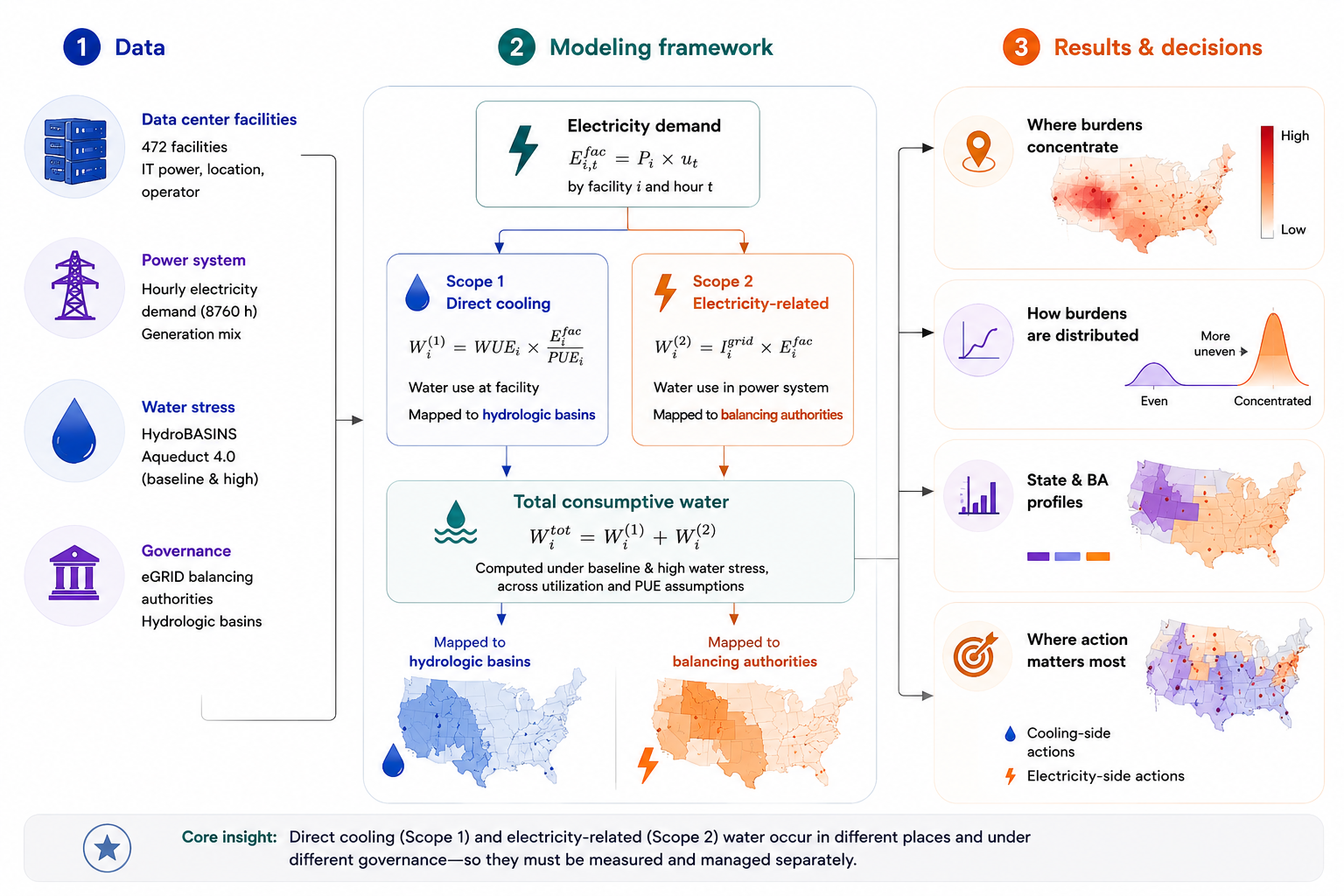}
\caption{\textbf{Study design.} Facility inventory, grid generation mix, and hydrologic stress data are linked to each hyperscale facility. Facility electricity demand is translated into two water pathways: Scope~1, consumed at the site and mapped to hydrologic basins, and Scope~2, consumed in the power system and mapped to balancing authorities. Pathway-separated estimates support hotspot maps, concentration analysis, and comparison maps of where different actions would reduce water use most.}
\label{fig:schematic}
\end{figure}

We developed a facility-resolved national framework for estimating annual operational water consumption from 472 U.S. hyperscale data centers across the two pathways (Fig.~\ref{fig:schematic}). Facility locations were linked to balancing authorities, hydrologic basins, and Aqueduct water-stress indicators through geospatial joins. Scope~1 was estimated using Water Usage Effectiveness (WUE), defined in ISO/IEC~30134-9 as litres of site water consumption per kWh of IT electricity \citep{iso30134-9}. Scope~2 was estimated by combining facility electricity demand with balancing-authority generation mixes from eGRID \citep{epaegrid} and technology-specific water-consumption factors from the literature \citep{macknick2012water,meldrum2013lifecycle}. Unless otherwise noted, results refer to the baseline scenario; efficiency, high-load, and no-hydro sensitivities are reported in the Supplementary Materials.

\subsection*{Electricity-related water dominates the national total}

The 472 facilities in the baseline inventory represent \num{20041}~MW of hyperscale nameplate capacity and approximately 116~TWh~yr$^{-1}$ of facility electricity demand. Their operational water consumption totals approximately 300~GL~yr$^{-1}$---nearly 300 billion litres per year---under the baseline scenario (range 205--451~GL~yr$^{-1}$ across the efficiency and high-load scenarios; table~S2). For reference, this baseline total is comparable to the annual residential water consumption of a metropolitan area of roughly 2--3 million people \citep{mayer1999residential}. Of the baseline total, approximately 74~GL~yr$^{-1}$ is direct cooling water (Scope~1) and approximately 226~GL~yr$^{-1}$ is electricity-related water (Scope~2). Scope~2 therefore contributes about 75\% of the national total, roughly three times the Scope~1 contribution.

An important methodological caveat applies to the Scope~2 estimate. We attribute reservoir evaporation to hydropower generation at 8.0~L~kWh$^{-1}$ in the baseline \citep{macknick2012water}, a convention that is methodologically contested \citep{mekonnen2012sustainability}. Under a no-hydro sensitivity that sets this factor to zero, Scope~2 drops from 226 to approximately 128~GL~yr$^{-1}$---a 43\% reduction---and the national total falls to approximately 202~GL~yr$^{-1}$ (table~S2). Even in this conservative case, Scope~2 remains the larger component (63\% of the total). We report both the baseline and no-hydro results throughout the Supplementary Materials so that readers can assess the influence of this single assumption.

As a partial external check, the three largest hyperscale operators (Google, Microsoft, and Meta) collectively reported approximately 25~GL of direct water consumption in their most recent sustainability disclosures \citep{google2024environmental,microsoft2024environmental,meta2024sustainability}. These operators represent a substantial but incomplete share of the facilities in our inventory. Scaling by their approximate share of the sample's total capacity yields a Scope~1 estimate broadly consistent with our baseline, although exact comparison is limited by differences in reporting scope and boundary definitions.

This national split is important, but it is not the main result. The more consequential finding is that the two pathways create different geographies of burden. Scope~1 is tied to the site, its water source, and the surrounding hydrologic context. Scope~2 is tied to the regional generation mix and the way new electricity demand is served. A single national footprint cannot show which institutions should act in which places.

\subsection*{The two pathways create different hotspots}

We mapped Scope~1 and Scope~2 on different spatial units because water is consumed in different places. Direct cooling water is consumed at or near the facility, so hydrologic basins are the appropriate geography for identifying where cooling burden overlaps hydrologic stress. Electricity-related water is consumed in the power system, often far from the facility, so balancing authorities---the operational units of the U.S. electricity grid---are the appropriate unit for electricity--water attribution.

\begin{figure}[t]
\centering
\includegraphics[width=0.95\textwidth]{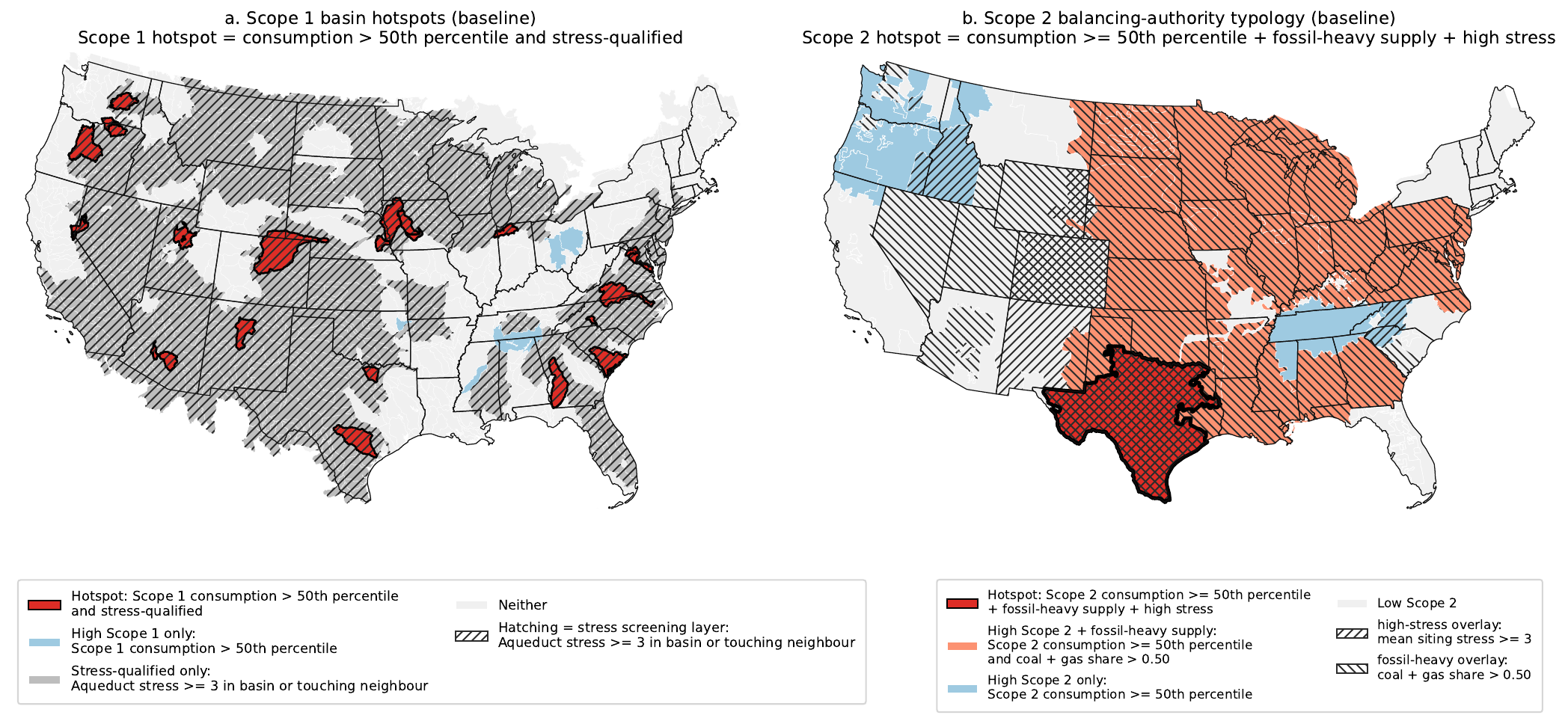}
\caption{\textbf{Direct cooling and electricity-related water produce different hotspot maps.} \textbf{A}, Scope~1 basin classes combine direct cooling-water burden (above-median among hosting basins) with basin water stress (Aqueduct score $\geq 3$ in the basin or a touching neighbor). \textbf{B}, Scope~2 balancing-authority classes combine electricity-related water burden (above-median among hosting BAs), fossil-heavy supply (coal + gas $>$50\%), and stress at hosted facility locations (MW-weighted mean $\geq 3$). The maps show that the places where cooling demand overlaps stressed hydrology are not the same as the places where data-center electricity demand is served by water-intensive power supply.}
\label{fig:hotspots}
\end{figure}

Figure~\ref{fig:hotspots} shows that these two hotspot geographies diverge. In the Scope~1 map, a hotspot is a basin where direct cooling-water consumption is above the median among hosting basins \emph{and} Aqueduct baseline water stress is high ($\geq 3$ on the 0--5 scale, classified by Aqueduct~4.0 as high to extremely high) in the basin or in a touching neighbor \citep{wriAqueduct,kuzma2023aqueduct40}. In the Scope~2 map, the most restrictive class identifies balancing authorities where electricity-related water is high, coal and gas provide more than half of annual generation, and hosted facilities are located in high-stress settings. Scope~1 hotspots highlight basins---primarily in the western and south-central U.S.---where data-center cooling demand overlaps stressed hydrology. Scope~2 hotspots highlight grid regions---often in the eastern U.S.---where large data-center loads are served by relatively water-intensive electricity. These are different problems for different decision makers.

\subsection*{Electricity-related water is more concentrated}

\begin{figure}[t]
\centering
\includegraphics[width=0.95\textwidth]{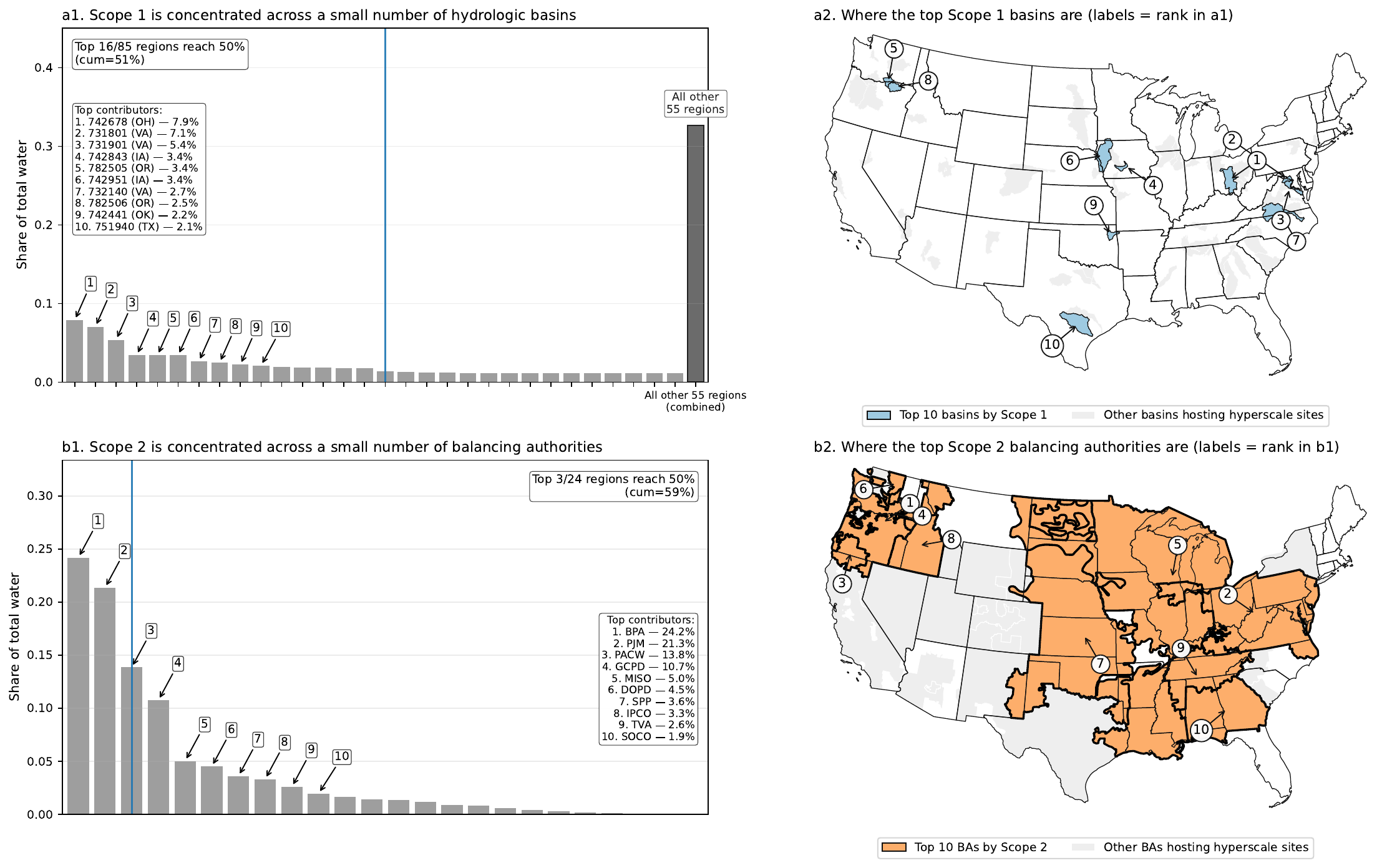}
\caption{\textbf{Electricity-related water is more spatially concentrated than direct cooling water.} \textbf{A}, Scope~1 concentration across hydrologic basins hosting hyperscale facilities. \textbf{B}, Scope~2 concentration across balancing authorities hosting hyperscale facilities. Bars rank regions by their share of total baseline pathway-specific water; maps show the leading regions. Scope~2 reaches half the national total in far fewer regions than Scope~1.}
\label{fig:concentration}
\end{figure}

The two pathways also differ in how unevenly their burdens are distributed across regions (Fig.~\ref{fig:concentration}). For Scope~1, the top 16 of 85 hosting basins account for 51\% of baseline Scope~1 water. For Scope~2, the top 3 of 24 hosting balancing authorities account for 59\% of baseline Scope~2 water. Thus, electricity-related water is dominated by a few grid regions, whereas direct cooling water is spread across a larger number of hydrologic basins.

This concentration has practical consequences. A highly concentrated electricity-side problem can in principle be addressed through a relatively small number of utility, procurement, and regional planning decisions. Cooling-water concerns require more local, basin-specific attention. The regions contributing most to Scope~1 are not the same as those contributing most to Scope~2, reinforcing the need to keep the pathways separate.

\subsection*{Different actions matter in different places}

\begin{figure}[t]
\centering
\includegraphics[width=0.95\textwidth]{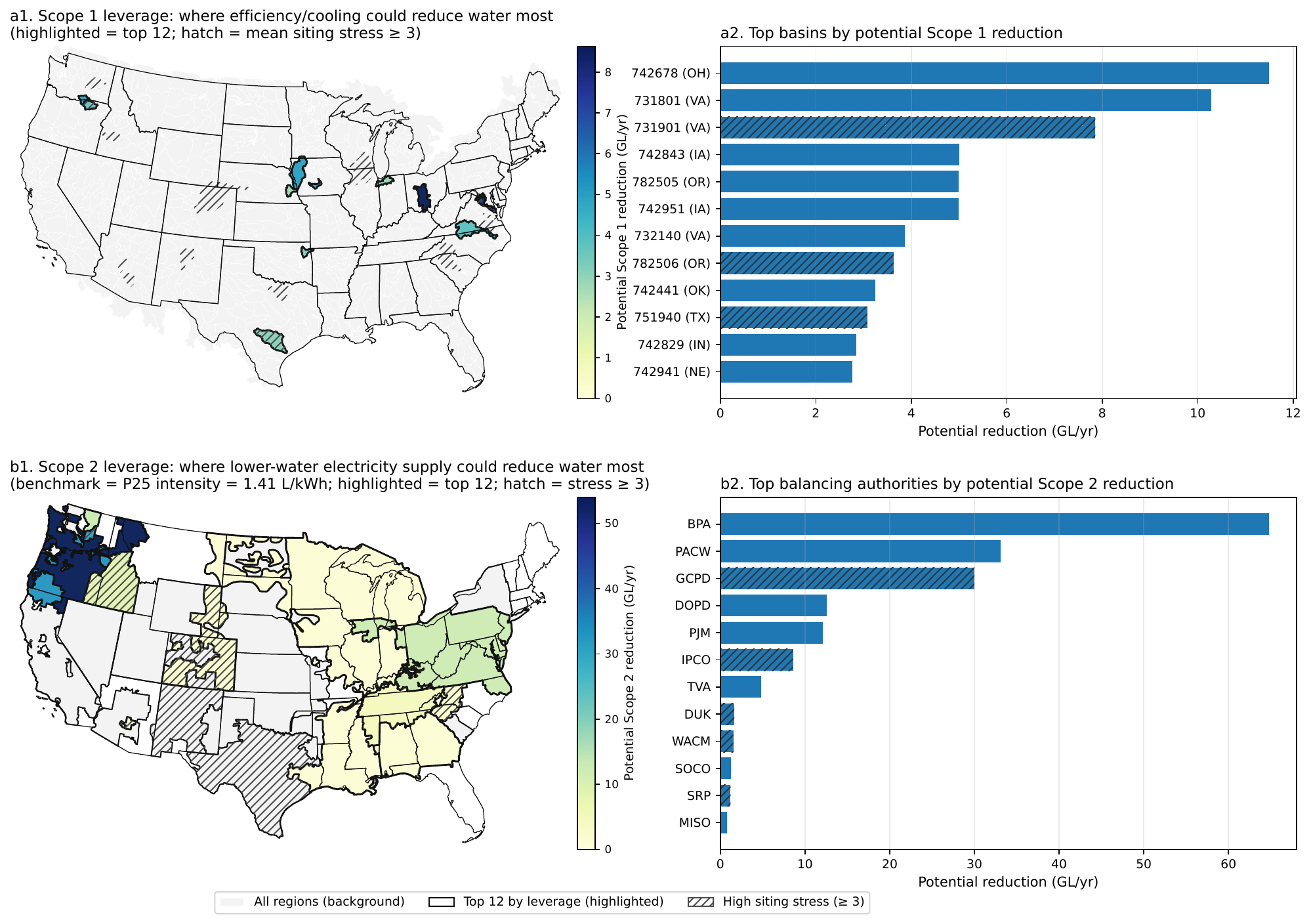}
\caption{\textbf{Different actions would reduce water use most in different places.} \textbf{A}, Potential Scope~1 reduction by hydrologic basin under a lower-water operating benchmark. Large values indicate both high current direct cooling water and a large gap to the benchmark. Hatching marks high water stress ($\geq 3$). \emph{Audience: local water managers, facility operators, permitting authorities. Action: less water-intensive cooling, reclaimed water, drought planning.} \textbf{B}, Potential Scope~2 reduction by balancing authority if grid water intensity were reduced to the 25th percentile among hosting BAs (\num{1.41}~L~kWh$^{-1}$). Large values indicate both high hosted load and a large grid-intensity gap. \emph{Audience: electricity regulators, utility planners, large purchasers. Action: less water-intensive generation, revised resource plans, expanded transmission.} The two maps identify different priority regions for cooling-side and electricity-side action.}
\label{fig:actions}
\end{figure}

The hotspot and concentration results show where burdens are large. We next asked where water use would fall most under simple lower-water comparison benchmarks (Fig.~\ref{fig:actions}). The mapped quantity is potential reduction in annual water consumption: the difference between the baseline estimate and the value that would remain if the region moved to the stated benchmark. These are comparison calculations, not forecasts or engineering-feasibility assessments.

For Scope~1, the benchmark is a lower-water operating case based on the efficiency scenario values of utilization, PUE, and WUE. Because these three parameters are applied uniformly across facilities, the geographic ranking of Scope~1 reduction potential is determined by existing capacity distribution---the same basins that appear large in Fig.~\ref{fig:hotspots}A also appear large here. The map is therefore most useful for quantifying the \emph{magnitude} of potential savings rather than for revealing new geography. A large basin value indicates high current cooling-water consumption and a large gap to that reference point. These basins are where local water managers, permitting authorities, and facility operators should examine cooling-system choices, water sourcing, reclaimed-water use, and drought planning.

For Scope~2, the benchmark reduces each balancing authority's grid water intensity to the 25th percentile among hosting balancing authorities (\num{1.41}~L~kWh$^{-1}$ in the baseline). Because grid water intensity varies substantially across BAs, the Scope~2 map adds genuine geographic information beyond the burden map: some BAs with large hosted loads already have low grid-water intensity and therefore show small reduction potential, whereas others with large loads and water-intensive supply show large potential. These are the grid regions where utility planners, public utility commissions, balancing authorities, and large electricity buyers should examine lower-water generation, procurement choices, resource plans, and transmission access.

The high-potential regions differ between panels. The places where cooling-side changes would save the most water are not the places where electricity-side changes would save the most water. This is the central action-oriented result: reducing data-center water use is not one problem with one solution, but two connected problems with different geographies and different decision makers.

\subsection*{Implications}

Three conclusions follow. First, data-center water governance should address both pathways, not only on-site cooling. Permitting and environmental review often focus on direct water use, but for many facilities the larger operational burden is in the electricity system. A state such as Virginia, the nation's largest data-center market, should not be evaluated only through local cooling permits when much of its data-center water burden is tied to electricity supply (Fig.~S5). Conversely, some western states with Scope~1-dominant profiles need locally tailored cooling and water-sourcing strategies.

Second, the largest reductions in electricity-related water are concentrated in a small number of grid regions. Because three balancing authorities account for more than half of Scope~2 water, decisions about generation mix, transmission, and procurement in these regions can disproportionately affect the national data-center water footprint. Many hyperscale operators have signed large renewable power-purchase agreements, but our location-based attribution reflects the physical water intensity of the grid as currently operated rather than contractual procurement claims; as these contracts are fulfilled with built generation, the location-based Scope~2 estimates will decline accordingly.

Third, cooling-water reductions require locally tailored strategies across more basins. Effective actions include less water-intensive cooling systems, reclaimed or non-potable water, and drought-contingency planning, but these choices must be adapted to local hydrologic conditions. The main value of pathway separation is therefore practical: it identifies where water is consumed, through which mechanism, and which decisions are most relevant to reducing it.

The analysis has limitations that should guide interpretation. It is annual-average and location-based; it does not resolve seasonal drought coincidence, dispatch-level electricity, exact source-water provenance, or facility-specific measured WUE. All facilities receive the same scenario-based WUE, so the spatial heterogeneity of Scope~1 reflects capacity distribution and location rather than observed differences in cooling technology across sites. The Scope~1 estimates should therefore be understood as showing where cooling-water \emph{exposure} is largest given current siting, not as measured consumption at individual facilities. Hydropower attribution remains contested and substantially affects the Scope~2 estimate, as described above. The boundary excludes embodied and supply-chain water. These limitations affect absolute totals more than the main geographic finding: cooling-side and electricity-side water concerns concentrate in different places, and this conclusion holds across all scenarios tested (Figs.~S1--S4).

\section*{References and Notes}
\bibliographystyle{unsrtnat}
\bibliography{bib}

\section*{Acknowledgments}
The authors thank colleagues for feedback on earlier versions.\\
\textbf{Funding:} Project Number R01MD016054; Project Number U24ES035309, with Michelle Bell and Nicole Deziel as PD/PIs; R01ES037156; and R01ES036731.\\
\textbf{Author contributions:} G.G. conceived and implemented the analysis. G.G. and F.D. designed the study, interpreted the results, and wrote the manuscript.\\
\textbf{Competing interests:} The authors declare no competing interests.\\
\textbf{Data and materials availability:}
All code necessary to reproduce the analysis workflow is available at
\texttt{https://github.com/gianguidi/hyperscale-water-geography}. The version
corresponding to this submission is archived at commit
\texttt{2b8672e8cef27165d3be5c2084946984d0e96c40}. Because the underlying facility-level dataset contains commercially sensitive information, raw facility identifiers, addresses, and coordinates are not publicly released. The repository includes a reproducibility notebook, environment files, configuration templates, and instructions for reproducing the workflow with authorized or synthetic inputs.

\section*{Supplementary Materials}
Materials and Methods\\
Supplementary Text\\
Figs.~S1 to S6\\
Tables~S1 to S4

\clearpage

% =========================================================
% SUPPLEMENTARY MATERIALS START HERE
% =========================================================

\setcounter{page}{1}
\setcounter{section}{0}
\setcounter{subsection}{0}
\setcounter{figure}{0}
\setcounter{table}{0}
\setcounter{equation}{0}

\renewcommand{\thesection}{S\arabic{section}}
\renewcommand{\thesubsection}{S\arabic{section}.\arabic{subsection}}
\renewcommand{\thefigure}{S\arabic{figure}}
\renewcommand{\thetable}{S\arabic{table}}
\renewcommand{\theequation}{S\arabic{equation}}

\begin{center}
    {\Large \ Supplementary Materials for}\\[0.75em]
    {\LARGE \ The Hidden Water Geography of U.S. Hyperscale Data Centers in the AI Era}\\[1em]
    {\large Gianluca Guidi and Francesca Dominici}
\end{center}

\bigskip

\tableofcontents
\clearpage

\section{Materials and Methods}

\subsection{Study design}
We estimate annual operational water consumption from U.S. hyperscale data centers using a facility-resolved framework that separates direct cooling water (Scope~1) from electricity-related water consumption (Scope~2). Scope~1 denotes water consumed directly at the facility for cooling. Scope~2 denotes water consumed off site to generate the electricity used by the facility. The core boundary includes operational water consumption only and excludes embodied and supply-chain water.

For each facility $i$,
\begin{equation}
W_i^{\mathrm{tot}} = W_i^{(1)} + W_i^{(2)}.
\end{equation}
When water-intensity factors are in L~kWh$^{-1}$ and electricity is in kWh, the combined expression is
\begin{equation}
W_i^{\mathrm{tot}} = 10^{-3}\left[
\underbrace{I_{r(i)}^{\mathrm{grid}} E_{i,\mathrm{kWh}}^{\mathrm{fac}}}_{\substack{\text{Scope 2:}\\\text{electricity-related water}}}
+
\underbrace{\mathrm{WUE}_i\frac{E_{i,\mathrm{kWh}}^{\mathrm{fac}}}{\mathrm{PUE}_i}}_{\substack{\text{Scope 1:}\\\text{direct cooling water}}}
\right],
\end{equation}
where $10^{-3}$ converts litres to cubic metres.

\subsection{Data sources}
Facility locations, identifiers, and power attributes come from a facility-level hyperscale inventory and electricity-attribution pipeline \citep{guidi2026assessingcarbonemissionsenergy}. The main analysis uses 472 U.S. hyperscale facilities in the 12-month study window ending in March 2026 that have current nameplate-power estimates and the spatial information needed for grid, basin, and water-stress joins.

Balancing-authority regions are assigned using point-in-polygon joins between facility coordinates and eGRID balancing-authority polygons \citep{epaegrid}. Generation shares by fuel or technology class are obtained from eGRID-linked summaries and normalized to sum to one. Hydrologic basins are assigned using HydroBASINS level-6 polygons \citep{lehner2013hydrobasins}. Baseline water-stress indicators are from Aqueduct Water Risk Atlas 4.0 \citep{wriAqueduct,kuzma2023aqueduct40}. Facilities without valid spatial assignments for a required layer are excluded from analyses that require that layer; missing grid-mix or stress attributes are not imputed.

\subsection{Facility electricity}
The facility attribute \texttt{current\_mw} is interpreted as nameplate facility-level power $P_i$ (MW). Annual facility electricity is
\begin{equation}
E_i^{\mathrm{fac}} = P_i \times 8760 \times u,
\end{equation}
where $u$ is an annual-average utilization factor. Because $P_i$ represents total facility power, PUE is not applied to scale facility electricity. PUE is used only to recover IT electricity for the WUE-based Scope~1 estimate. Electricity is converted to kWh as
\begin{equation}
E_{i,\mathrm{kWh}}^{\mathrm{fac}} = 1000 E_i^{\mathrm{fac}}.
\end{equation}

\subsection{Scope~1: direct cooling water}
Water Usage Effectiveness (WUE), defined in ISO/IEC~30134-9, is litres of site water use per kWh of IT electricity \citep{iso30134-9}. IT electricity is recovered as
\begin{equation}
E_{i,\mathrm{kWh}}^{\mathrm{IT}} = \frac{E_{i,\mathrm{kWh}}^{\mathrm{fac}}}{\mathrm{PUE}_i},
\end{equation}
and Scope~1 water consumption is
\begin{equation}
W_i^{(1)} = 10^{-3}\,\mathrm{WUE}_i\,E_{i,\mathrm{kWh}}^{\mathrm{IT}}.
\end{equation}
Facility-specific WUE is not consistently disclosed, so the analysis uses scenario windows informed by the literature \citep{mytton2021datacentre,lbnl2024dc,li2023thirsty,shehabi2016united}.

\subsection{Scope~2: electricity-related water}
Scope~2 water consumption is
\begin{equation}
W_i^{(2)} = 10^{-3} I_{r(i)}^{\mathrm{grid}} E_{i,\mathrm{kWh}}^{\mathrm{fac}},
\end{equation}
where $I_r^{\mathrm{grid}}$ is the balancing-authority water-consumption intensity. This intensity is calculated from generation shares and operational water-consumption factors:
\begin{equation}
I_r^{\mathrm{grid}} = \sum_{f\in\mathcal{F}} s_{r,f}w_f,
\end{equation}
where $s_{r,f}$ is the share of generation in region $r$ from fuel or technology class $f$, and $w_f$ is the corresponding operational water-consumption factor \citep{macknick2012water,meldrum2013lifecycle}. This is a location-based attribution: facilities inherit the average physical water intensity of the regional electricity system associated with their location, not contractual procurement claims.

\subsection{Water stress and aggregation}
Each facility is assigned an Aqueduct baseline water-stress score $S\in[0,5]$ and classified as high stress if $S\ge3$ \citep{wriAqueduct,kuzma2023aqueduct40}. Facility-level estimates are aggregated to states, balancing authorities, and hydrologic basins. For balancing-authority stress overlays, we use an MW-weighted mean siting-stress metric:
\begin{equation}
\bar S_r^{\mathrm{mw}} = \frac{\sum_{i:r(i)=r} P_iS_i}{\sum_{i:r(i)=r}P_i}.
\end{equation}

\subsection{Hotspot definitions}
For Scope~1, a basin is high burden if its total Scope~1 water is above the median among basins hosting hyperscale facilities. A basin is stress-qualified if Aqueduct baseline stress is $\ge3$ in the basin or in a touching neighbor. Scope~1 hotspots satisfy both conditions.

For Scope~2, a balancing authority is high burden if its total Scope~2 water is at or above the median among hosting balancing authorities. Fossil-heavy supply is defined as coal plus gas generation share greater than 0.5. High stress is defined as $\bar S_r^{\mathrm{mw}}\ge3$. The most restrictive Scope~2 class satisfies all three conditions: high Scope~2 burden, fossil-heavy supply, and high stress.

\subsection{Comparison maps in main-text Fig.~4}
For Scope~1, the lower-water comparison value is obtained by scaling baseline Scope~1 according to scenario assumptions:
\begin{equation}
W_{i,\mathrm{bench}}^{(1)} = W_{i,\mathrm{base}}^{(1)} \frac{u_{\mathrm{bench}}}{u_{\mathrm{base}}}\frac{\mathrm{WUE}_{\mathrm{bench}}/\mathrm{PUE}_{\mathrm{bench}}}{\mathrm{WUE}_{\mathrm{base}}/\mathrm{PUE}_{\mathrm{base}}}.
\end{equation}
Basin-level potential Scope~1 savings are the sum of facility-level differences within each basin.

For Scope~2, the comparison sets each balancing authority's grid water intensity to the minimum of its observed value and the 25th percentile among hosting balancing authorities:
\begin{equation}
I_r^{\mathrm{target}} = \min\{I_r^{\mathrm{grid}},Q_{0.25}(I_r^{\mathrm{grid}})\}.
\end{equation}
Potential Scope~2 savings are
\begin{equation}
\Delta W_r^{(2)} = 10^{-3}\left(I_r^{\mathrm{grid}}-I_r^{\mathrm{target}}\right)E_{r,\mathrm{kWh}}^{\mathrm{fac}},
\end{equation}
where $E_{r,\mathrm{kWh}}^{\mathrm{fac}}$ is total annual facility electricity hosted in balancing authority $r$. These benchmarks rank where reductions would be largest under stated assumptions. They are not forecasts, cost curves, or feasibility assessments.

\subsection{Scenario windows}
The baseline scenario is the central parameter set used in the main text. We also evaluate an efficiency scenario, a high-load scenario, and a no-hydro sensitivity. Under the facility-load interpretation, Scope~2 scales linearly with $u$, whereas Scope~1 scales with $u\times(\mathrm{WUE}/\mathrm{PUE})$. In the no-hydro sensitivity, the hydropower water factor is set to zero to bound the influence of reservoir-evaporation attribution.

\begin{table}[t]
\centering
\small
\setlength{\tabcolsep}{6pt}
\renewcommand{\arraystretch}{1.12}
\begin{tabular}{@{}p{0.20\linewidth}p{0.30\linewidth}p{0.22\linewidth}p{0.22\linewidth}@{}}
\toprule
\textbf{Quantity} & \textbf{Definition / units} & \textbf{Values} & \textbf{Notes} \\
\midrule
$u$ & Annual-average utilization & 0.55 / \textbf{0.66} / 0.85 & Efficiency / baseline / high-load \\
PUE & Facility/IT electricity & 1.15 / \textbf{1.25} / 1.40 & Used only to recover IT electricity \\
WUE & L~kWh$^{-1}_{\mathrm{IT}}$ & 0.2 / \textbf{0.8} / 1.5 & Direct cooling-water intensity \\
$w_f$ & L~kWh$^{-1}$ by generation class & Coal 1.9; gas 0.7; nuclear 2.5; hydro 8.0; wind 0.0; solar 0.1; other 1.0 & Baseline factor set \\
High stress & Aqueduct score & $S\ge3$ & High to extremely high stress \\
Scope~2 benchmark & Grid water-intensity target & $Q_{0.25}(I_r^{\mathrm{grid}})$ & Lower-water comparator \\
\bottomrule
\end{tabular}
\caption{Key parameters and scenario values. Baseline values are shown in bold.}
\label{tab:params_supp}
\end{table}

\section{Supplementary Text}

\subsection{National scenario sensitivity}
Table~\ref{tab:scenario_summary_supp} reports national totals across the scenario set. Absolute totals vary substantially, as expected from the scaling equations. Total operational water ranges from 204.91~GL~yr$^{-1}$ in the efficiency scenario to 450.60~GL~yr$^{-1}$ in the high-load scenario. The no-hydro sensitivity lowers Scope~2 from 225.74 to 128.35~GL~yr$^{-1}$. Scope~2 remains the larger component in all scenarios, although the difference narrows when hydropower attribution is removed.

\begin{table}[t]
\centering
\small
\begin{tabular}{lrrrrrrr}
\toprule
\textbf{Scenario} & \textbf{Electricity} & \textbf{Scope~1} & \textbf{Scope~2} & \textbf{Total} & \textbf{Scope~1} & \textbf{Scope~2} & \textbf{Intensity} \\
 & \textbf{(TWh~yr$^{-1}$)} & \textbf{(GL~yr$^{-1}$)} & \textbf{(GL~yr$^{-1}$)} & \textbf{(GL~yr$^{-1}$)} & \textbf{(\%)} & \textbf{(\%)} & \textbf{(L~kWh$^{-1}$)} \\
\midrule
Efficiency & 96.56 & 16.79 & 188.11 & 204.91 & 8.20 & 91.80 & 2.12 \\
Baseline & 115.87 & 74.16 & 225.74 & 299.89 & 24.73 & 75.27 & 2.59 \\
High-load & 149.22 & 159.88 & 290.72 & 450.60 & 35.48 & 64.52 & 3.02 \\
No-hydro & 115.87 & 74.16 & 128.35 & 202.50 & 36.62 & 63.38 & 1.75 \\
\bottomrule
\end{tabular}
\caption{National operational water summary across scenarios. Totals may not sum exactly because of rounding.}
\label{tab:scenario_summary_supp}
\end{table}

\subsection{Robustness of hotspot geography}
\begin{figure}[p]
\centering
\begin{minipage}{0.49\textwidth}
\centering
\includegraphics[width=\textwidth]{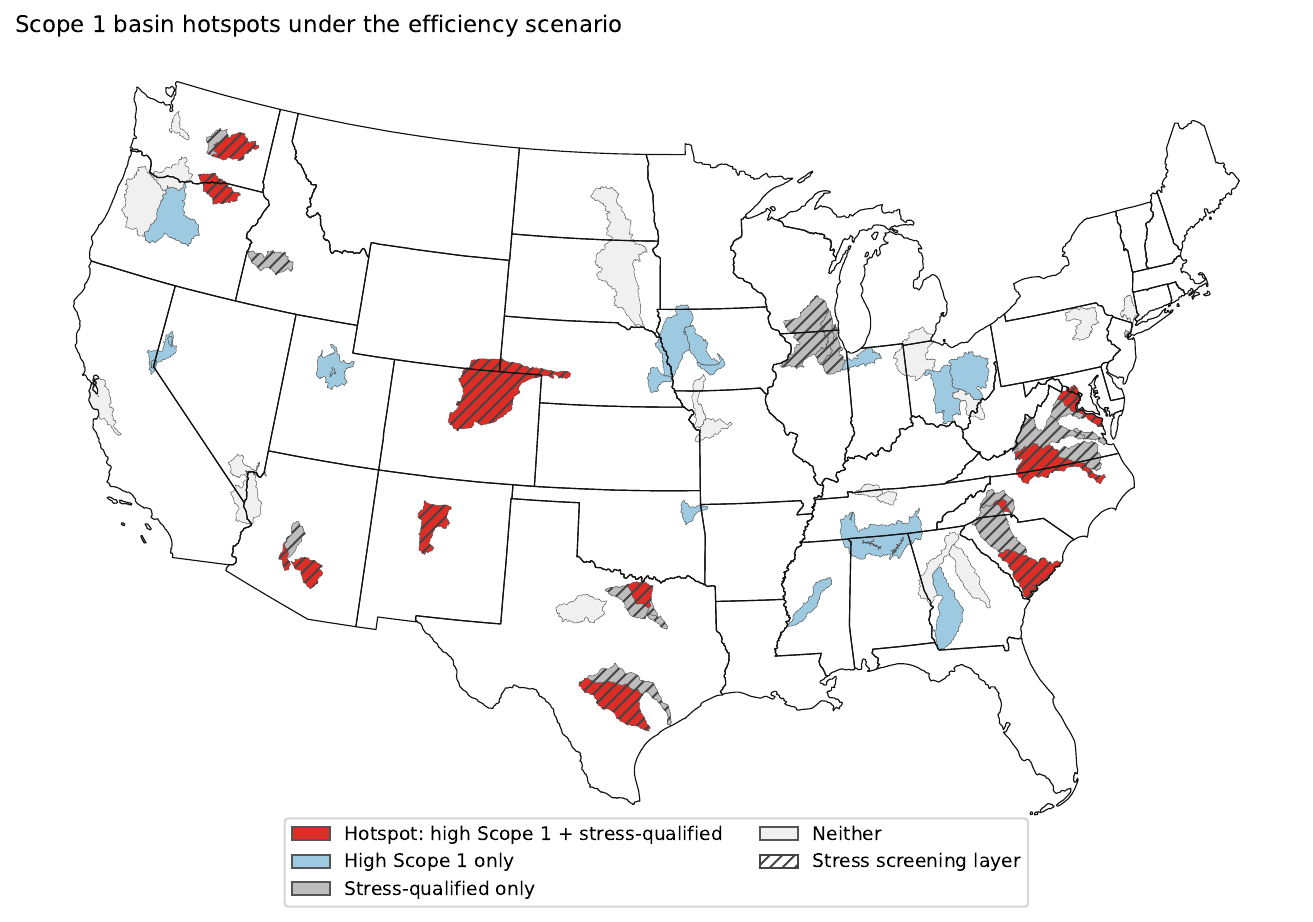}\\
\small \textbf{A}, Efficiency
\end{minipage}\hfill
\begin{minipage}{0.49\textwidth}
\centering
\includegraphics[width=\textwidth]{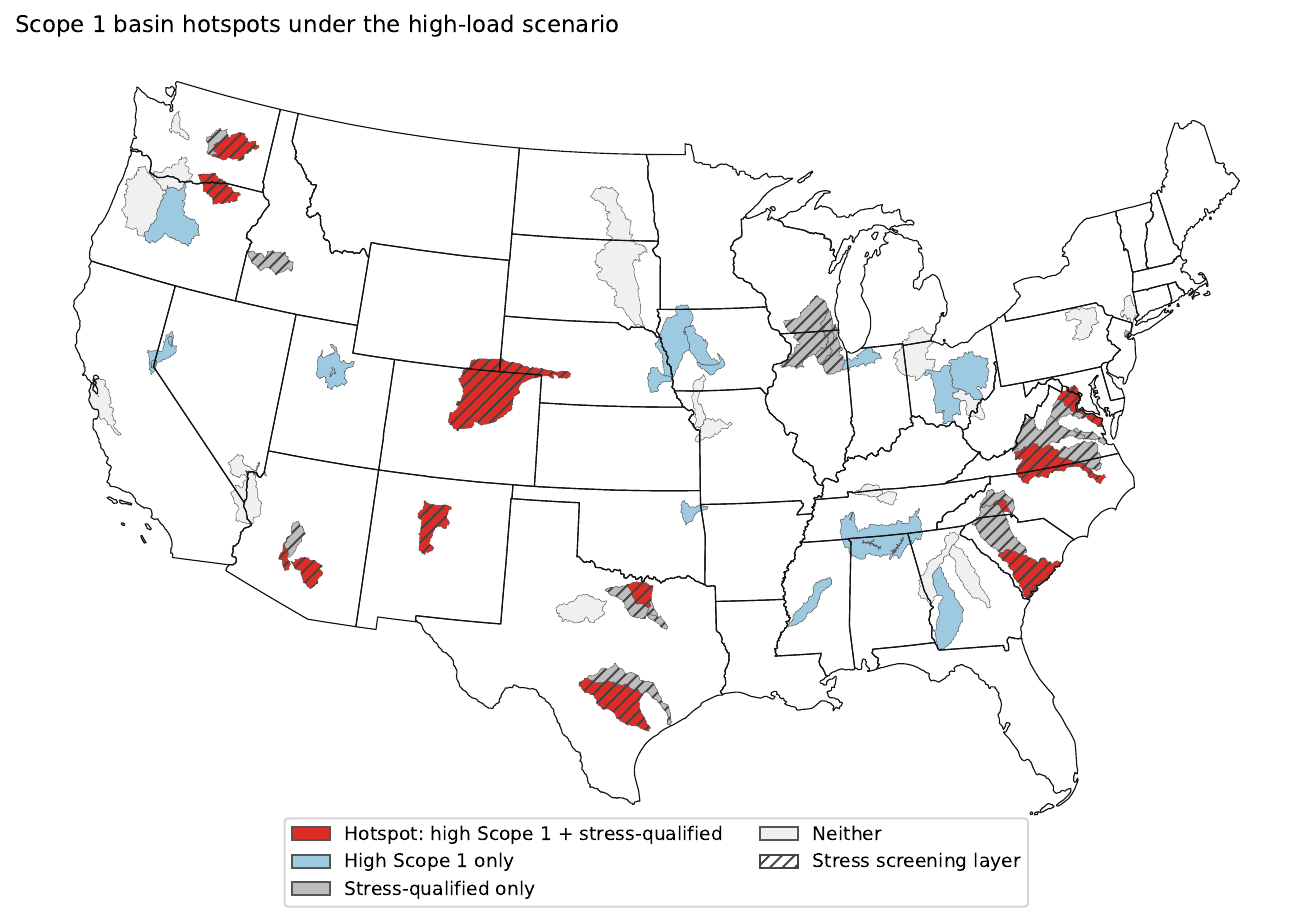}\\
\small \textbf{B}, High-load
\end{minipage}

\vspace{0.8em}

\begin{minipage}{0.72\textwidth}
\centering
\includegraphics[width=\textwidth]{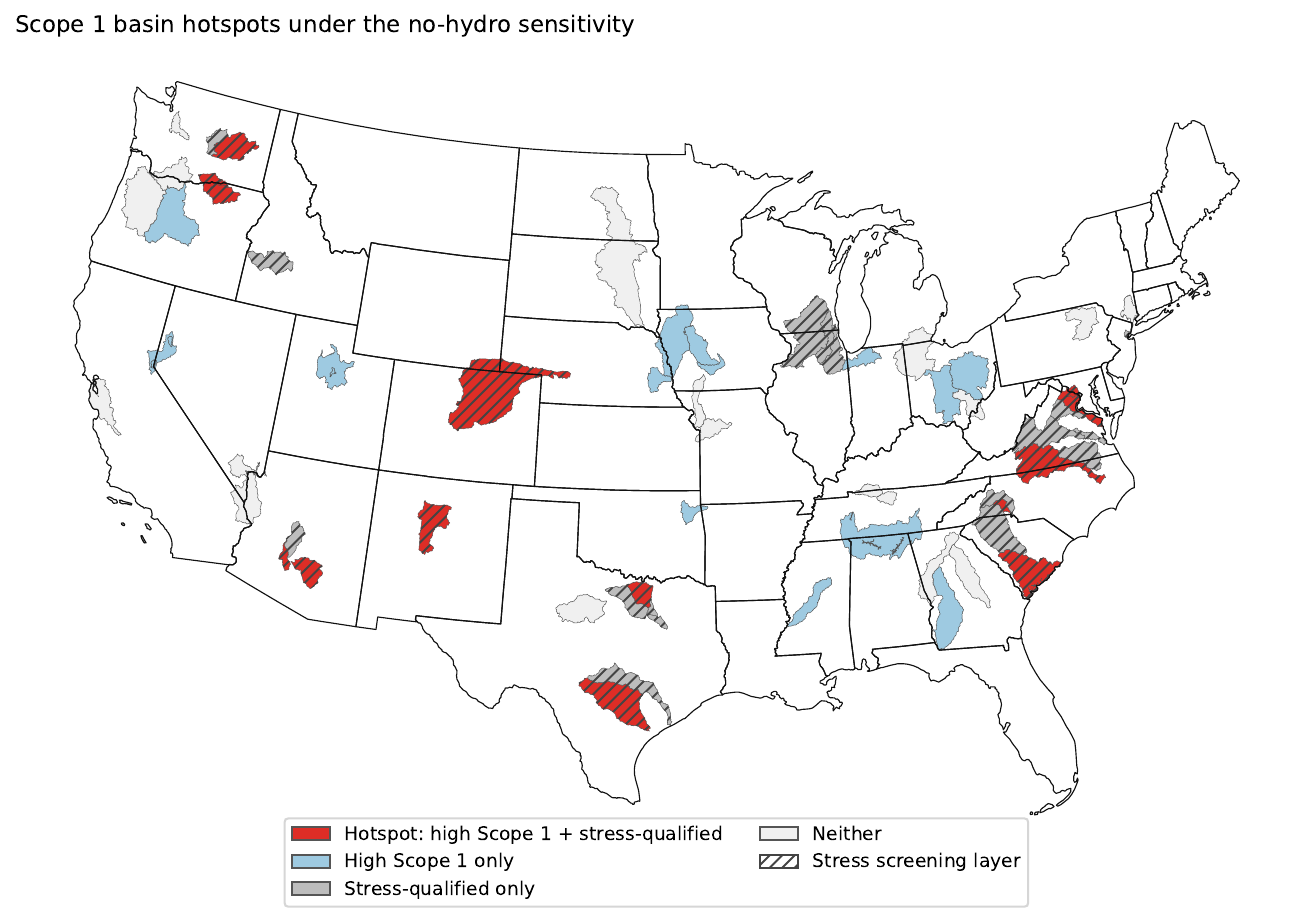}\\
\small \textbf{C}, No-hydro
\end{minipage}
\caption{\textbf{Scope~1 basin hotspot robustness.} Basins are classified using the same rule as in main-text Fig.~2A. The hotspot geography is visually unchanged across scenarios.}
\label{fig:s1_robustness_triptych}
\end{figure}

\begin{figure}[p]
\centering
\begin{minipage}{0.49\textwidth}
\centering
\includegraphics[width=\textwidth]{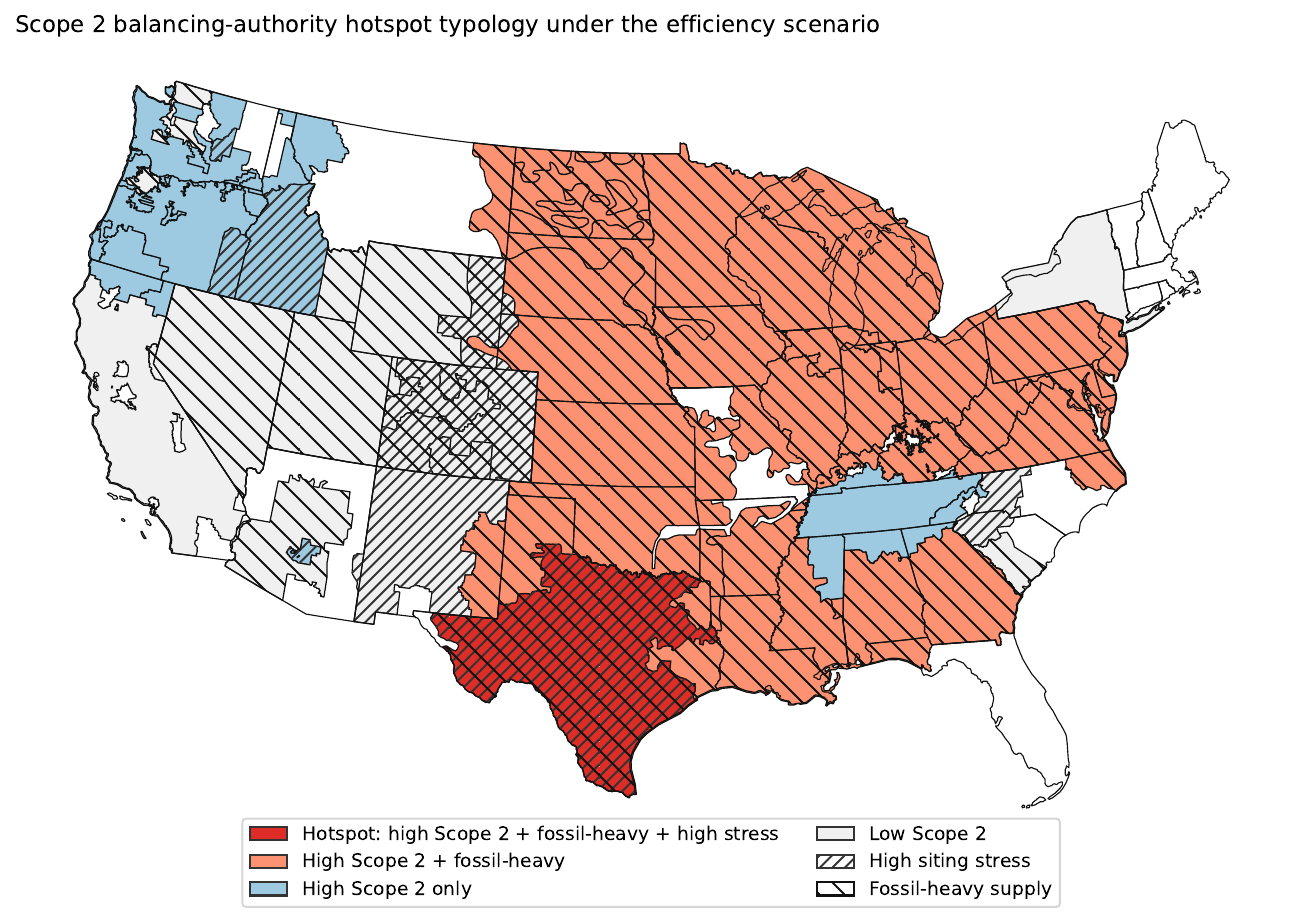}\\
\small \textbf{A}, Efficiency
\end{minipage}\hfill
\begin{minipage}{0.49\textwidth}
\centering
\includegraphics[width=\textwidth]{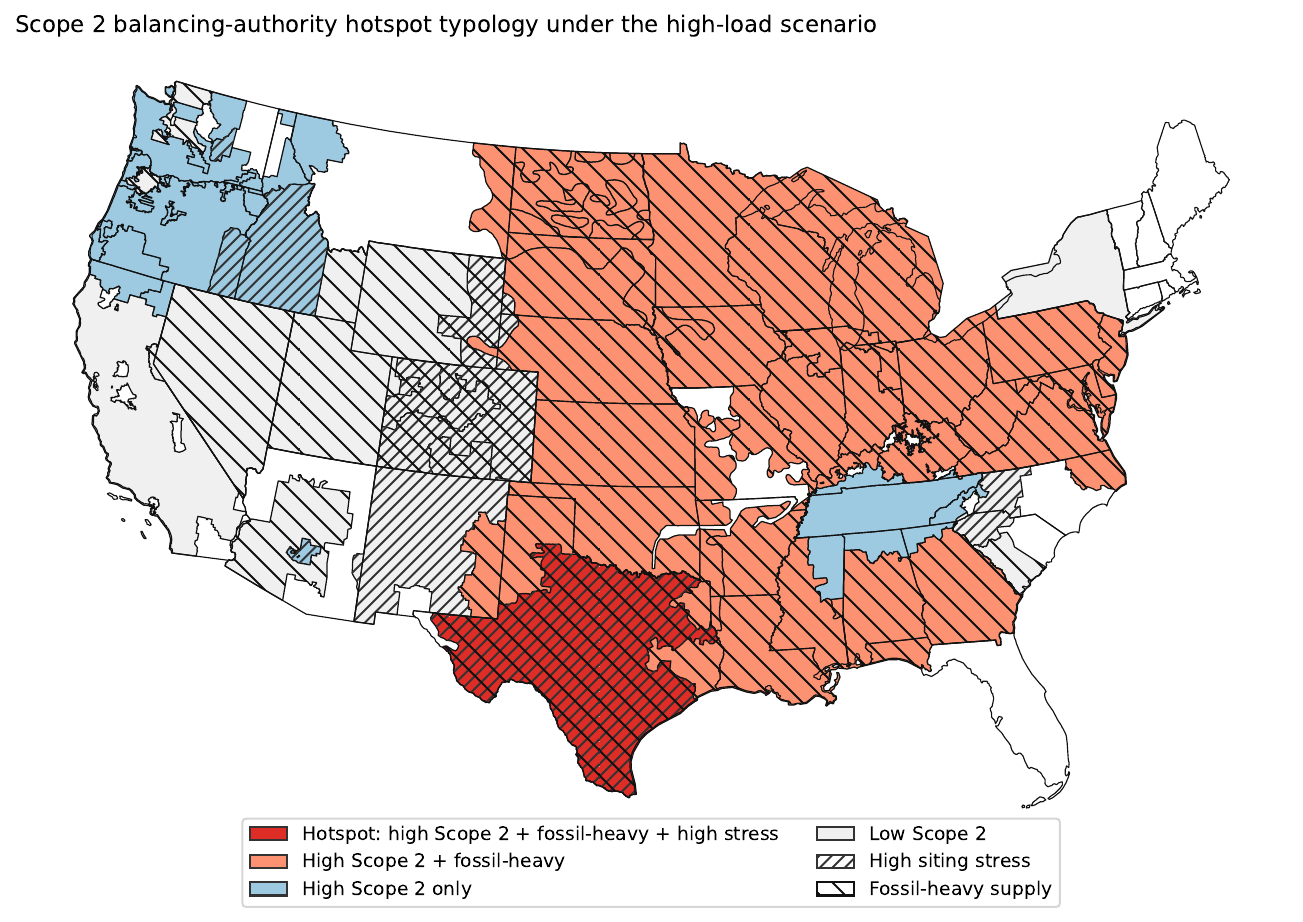}\\
\small \textbf{B}, High-load
\end{minipage}

\vspace{0.8em}

\begin{minipage}{0.72\textwidth}
\centering
\includegraphics[width=\textwidth]{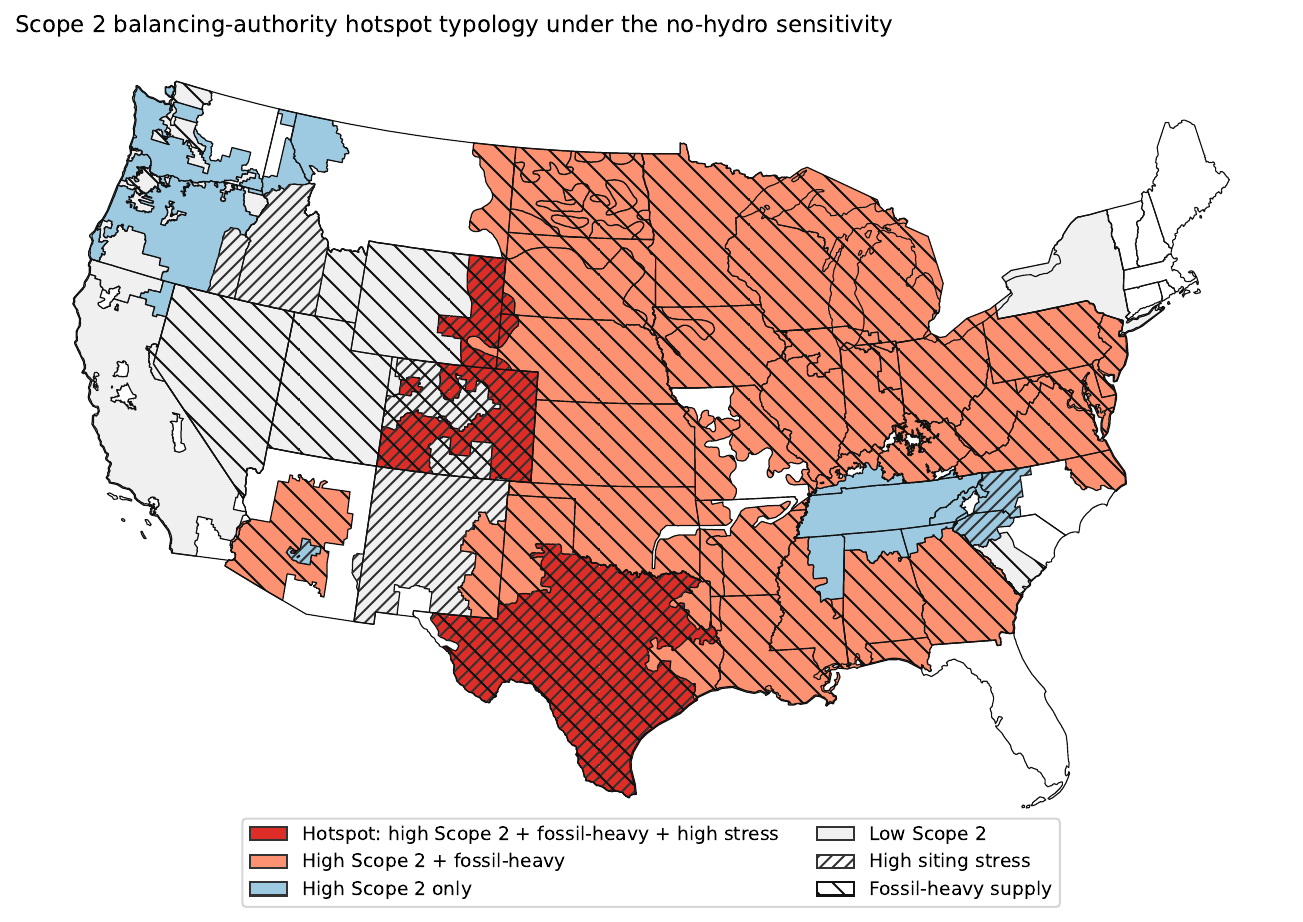}\\
\small \textbf{C}, No-hydro
\end{minipage}
\caption{\textbf{Scope~2 balancing-authority hotspot robustness.} Balancing authorities are classified using the same typology as in main-text Fig.~2B. Operating assumptions primarily rescale Scope~2 burden, whereas the no-hydro sensitivity changes some hydro-heavy western classifications.}
\label{fig:s2_robustness_triptych}
\end{figure}

Figures~\ref{fig:s1_robustness_triptych} and \ref{fig:s2_robustness_triptych} show that the main hotspot geographies are stable to operating assumptions. Scope~1 hotspot geography is visually unchanged across the efficiency, high-load, and no-hydro cases. Scope~2 geography is stable across the efficiency and high-load cases, whereas the no-hydro sensitivity changes the classification of some hydro-heavy western balancing authorities without eliminating the broader geography of electricity-related concern.

\subsection{Concentration and rank stability}
\begin{figure}[p]
\centering
\includegraphics[width=0.92\textwidth]{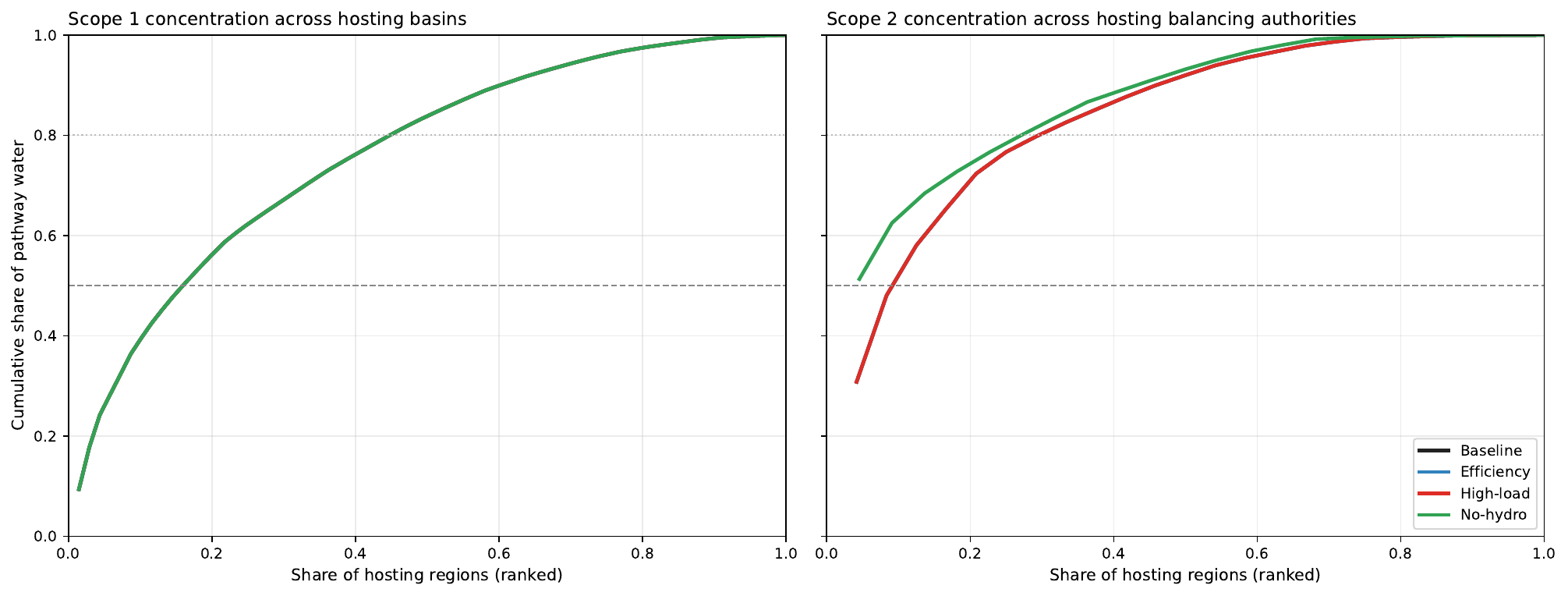}
\caption{\textbf{Concentration robustness across scenarios.} Cumulative concentration curves for Scope~1 across basins and Scope~2 across balancing authorities are shown for all scenarios. Scope~2 remains more spatially concentrated than Scope~1.}
\label{fig:concentration_robustness_supp}
\end{figure}

\begin{figure}[p]
\centering
\includegraphics[width=0.95\textwidth]{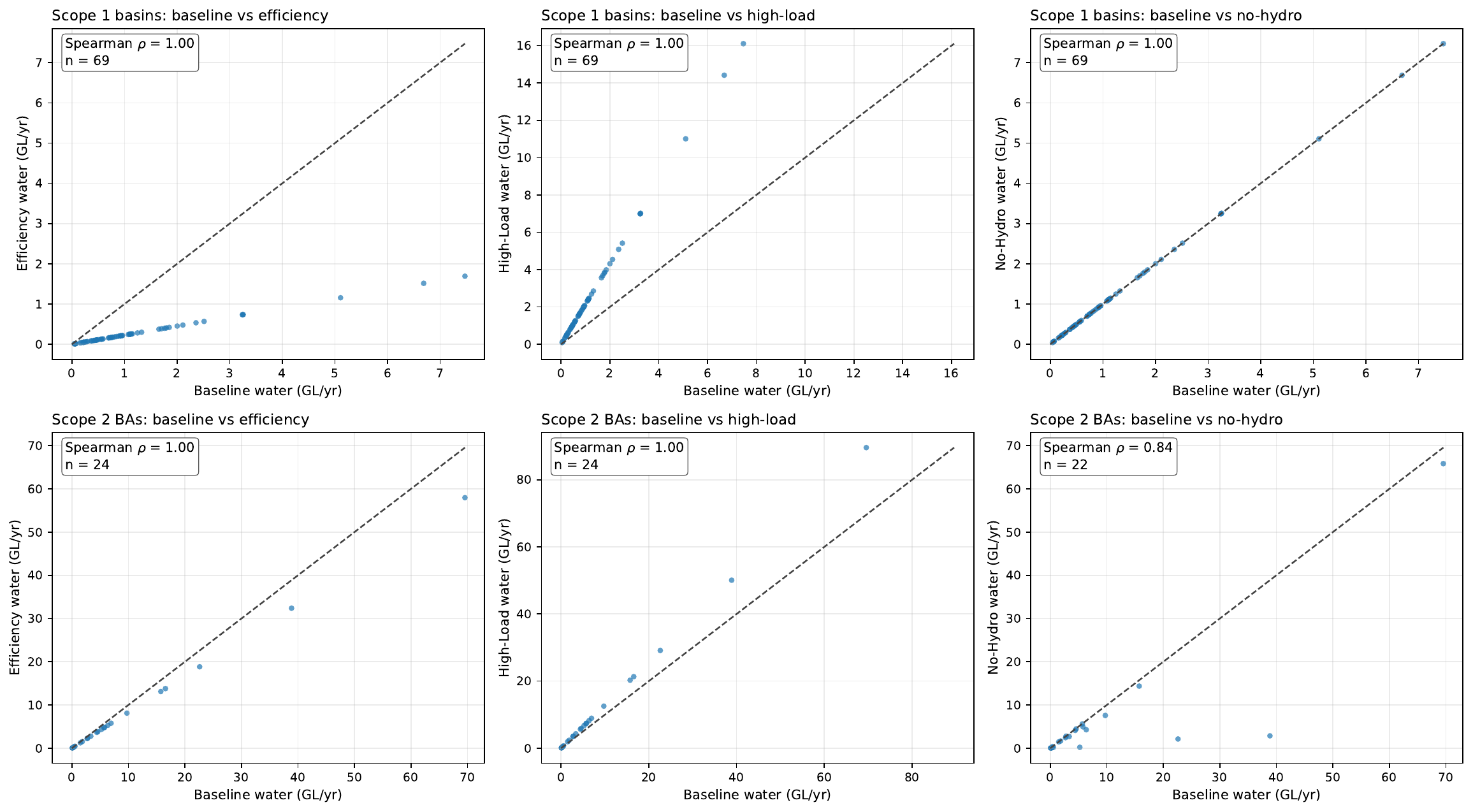}
\caption{\textbf{Rank stability across scenarios.} Each panel compares baseline regional burden with the corresponding burden under an alternative scenario. Spearman correlations show that operating assumptions change absolute totals more than regional rank ordering.}
\label{fig:rank_stability_supp}
\end{figure}

Figures~\ref{fig:concentration_robustness_supp} and \ref{fig:rank_stability_supp} show that the concentration result and leading-region ranks are robust. Cumulative concentration curves remain close across scenarios. Scope~1 basin rankings are invariant across scenarios, and Scope~2 balancing-authority rankings are invariant across efficiency and high-load cases and only partially altered under no-hydro.

\subsection{State and balancing-authority profiles}
\begin{figure}[p]
\centering
\includegraphics[width=0.98\textwidth]{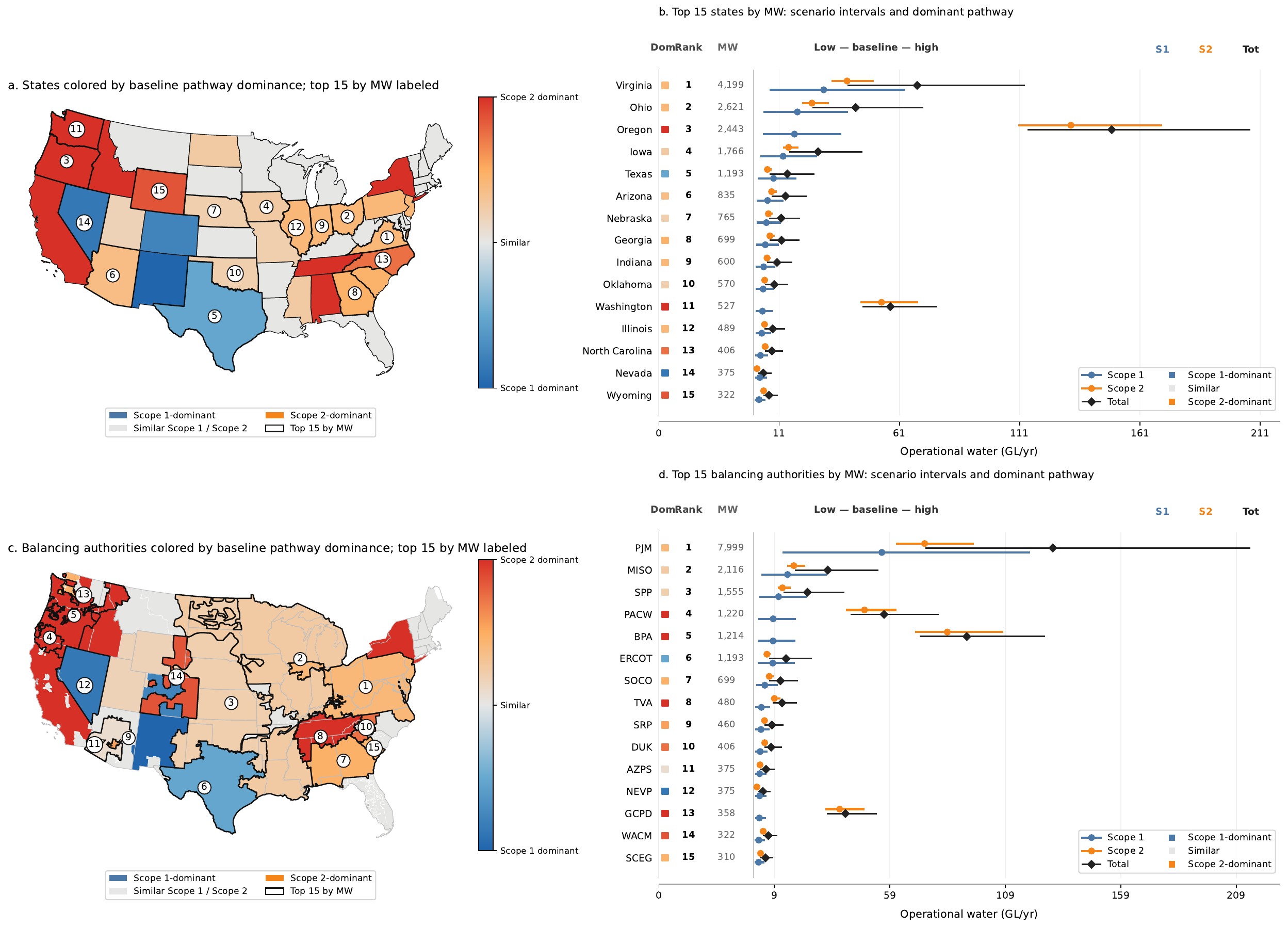}
\caption{\textbf{State and balancing-authority pathway profiles.} States and balancing authorities are colored by which pathway contributes more to baseline operational water. Ranked panels show low, baseline, and high scenario intervals for the 15 states and 15 balancing authorities with the most installed hyperscale capacity.}
\label{fig:state_ba_profiles_supp}
\end{figure}

Figure~\ref{fig:state_ba_profiles_supp} translates the pathway results into states and balancing authorities, units familiar to many decision makers. States are included because siting, permitting, and utility planning are often discussed at that scale. Balancing authorities are included because they are the operational unit most relevant for electricity-related water. Virginia illustrates a Scope~2-dominant state, whereas Texas and Nevada illustrate Scope~1-dominant or more cooling-focused profiles in the baseline scenario.

\subsection{Bivariate diagnostic for Scope~2}
\begin{figure}[p]
\centering
\begin{minipage}{0.49\textwidth}
\centering
\includegraphics[width=\textwidth]{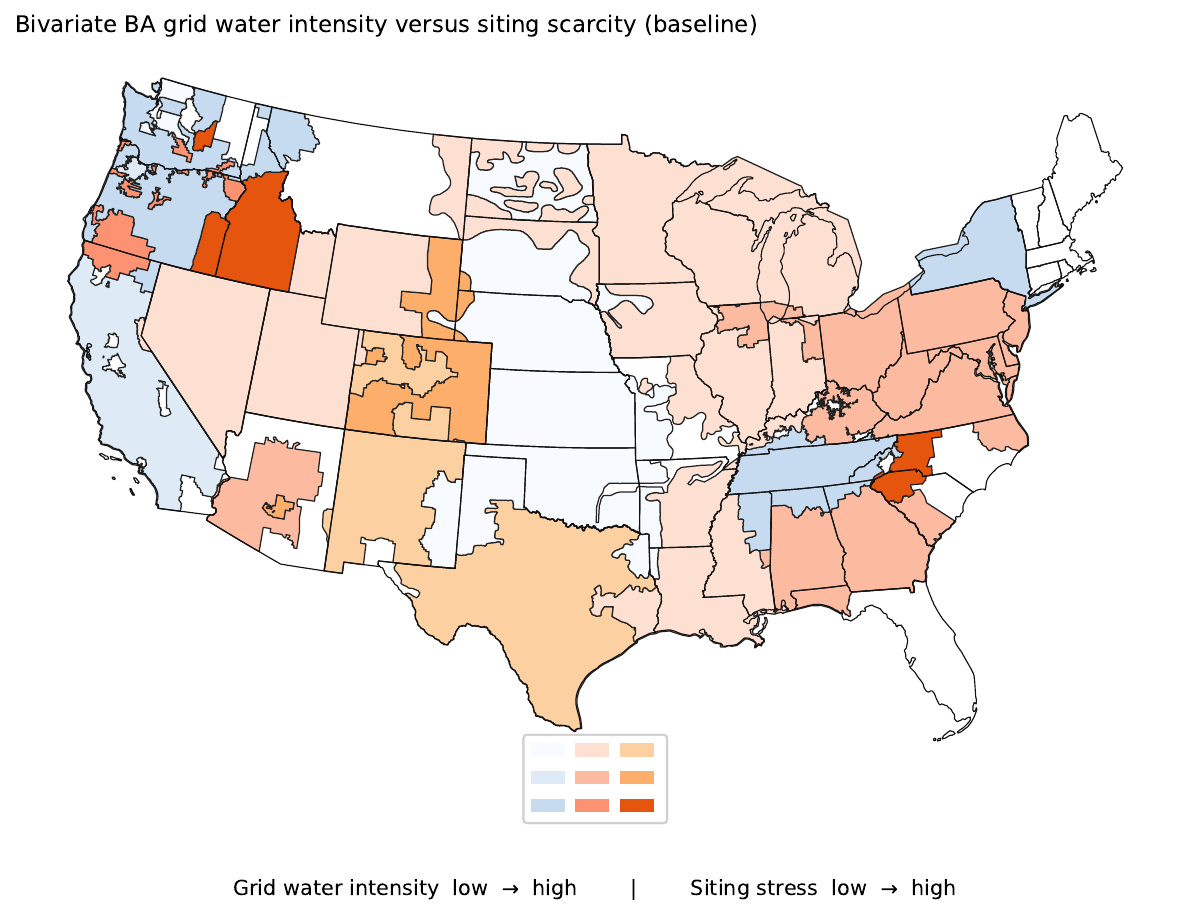}\\
\small \textbf{A}, Baseline
\end{minipage}\hfill
\begin{minipage}{0.49\textwidth}
\centering
\includegraphics[width=\textwidth]{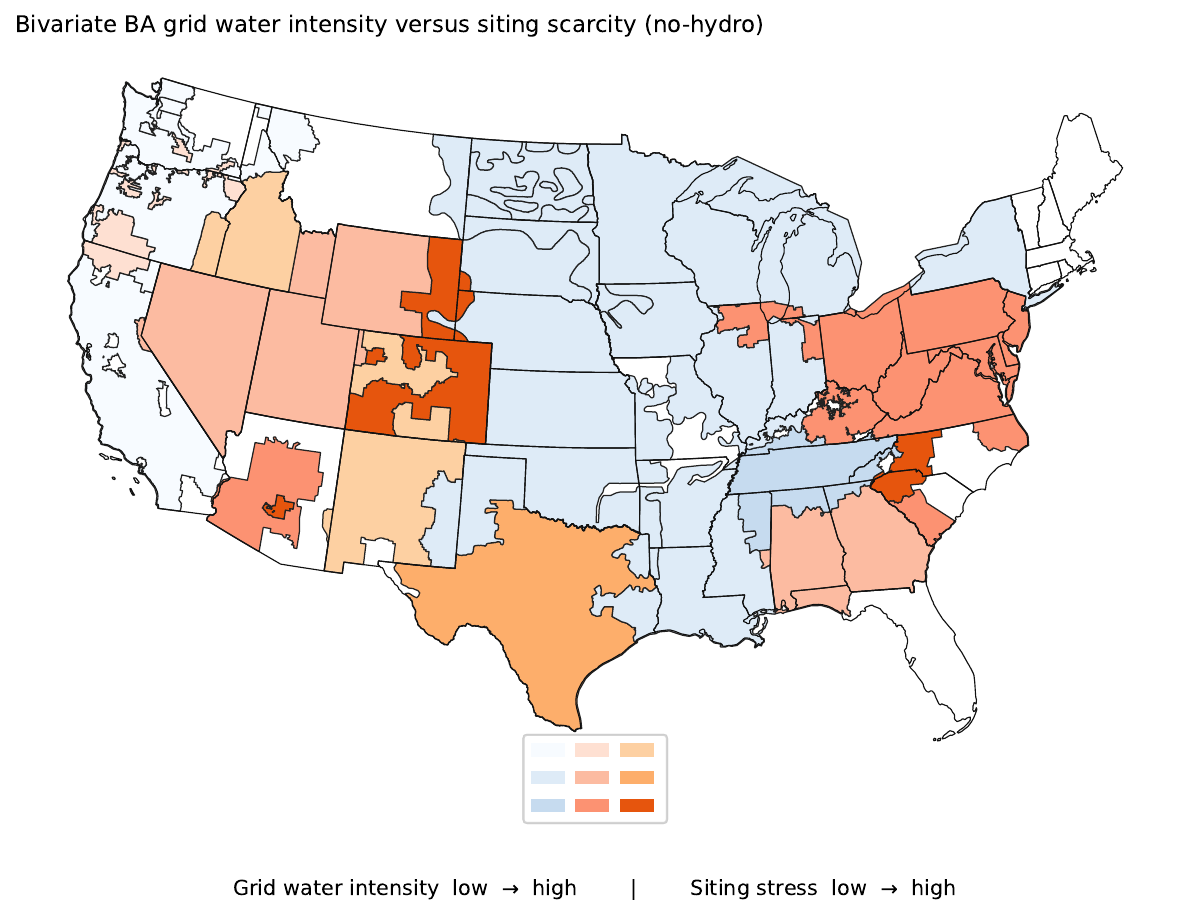}\\
\small \textbf{B}, No-hydro
\end{minipage}
\caption{\textbf{Bivariate balancing-authority grid water intensity versus siting stress.} Warmer upper-right classes indicate the joint occurrence of high grid water intensity and high siting stress. Excluding hydropower shifts some hydro-heavy western balancing authorities into lower intensity classes, while many elevated regions outside the West remain.}
\label{fig:bivariate_baseline_nohydro_supp}
\end{figure}

Figure~\ref{fig:bivariate_baseline_nohydro_supp} combines grid water intensity with MW-weighted mean siting stress. It is not a map of total burden. Instead, it shows where relatively water-intensive electricity supply and stressed hosting exposure co-occur. The baseline and no-hydro panels isolate the role of hydropower attribution.

\section{Supplementary Tables}

\begin{table}[p]
\centering
\scriptsize
\setlength{\tabcolsep}{4pt}
\begin{adjustbox}{max width=\textwidth}
\begin{tabular}{l r l l l l}
\toprule
State & MW & Scope 1 (GL/yr) & Scope 2 (GL/yr) & Total (GL/yr) & Total (ML/MW-yr) \\
\midrule
Virginia & 4,199 & \scen{6.6}{29.1}{62.8} & \scen{32.4}{38.9}{50.0} & \scen{39.0}{68.0}{112.8} & 9 / \textbf{16} / 27 \\
Ohio & 2,621 & \scen{4.1}{18.2}{39.2} & \scen{20.2}{24.3}{31.2} & \scen{24.3}{42.4}{70.4} & 9 / \textbf{16} / 27 \\
Oregon & 2,443 & \scen{3.8}{16.9}{36.5} & \scen{109.9}{131.9}{169.9} & \scen{113.8}{148.9}{206.4} & 47 / \textbf{61} / 84 \\
Iowa & 1,766 & \scen{2.8}{12.3}{26.4} & \scen{12.1}{14.5}{18.7} & \scen{14.9}{26.8}{45.1} & 8 / \textbf{15} / 26 \\
Texas & 1,193 & \scen{1.9}{8.3}{17.8} & \scen{4.8}{5.7}{7.4} & \scen{6.7}{14.0}{25.2} & 6 / \textbf{12} / 21 \\
Arizona & 835 & \scen{1.3}{5.8}{12.5} & \scen{6.2}{7.5}{9.6} & \scen{7.5}{13.3}{22.1} & 9 / \textbf{16} / 26 \\
Nebraska & 765 & \scen{1.2}{5.3}{11.4} & \scen{5.1}{6.1}{7.9} & \scen{6.3}{11.4}{19.4} & 8 / \textbf{15} / 25 \\
Georgia & 699 & \scen{1.1}{4.8}{10.5} & \scen{5.6}{6.8}{8.7} & \scen{6.7}{11.6}{19.2} & 10 / \textbf{17} / 27 \\
Indiana & 600 & \scen{0.9}{4.2}{9.0} & \scen{4.6}{5.6}{7.2} & \scen{5.6}{9.7}{16.1} & 9 / \textbf{16} / 27 \\
Oklahoma & 570 & \scen{0.9}{4.0}{8.5} & \scen{3.8}{4.6}{5.9} & \scen{4.7}{8.5}{14.4} & 8 / \textbf{15} / 25 \\
Washington & 527 & \scen{0.8}{3.7}{7.9} & \scen{44.3}{53.2}{68.5} & \scen{45.1}{56.8}{76.4} & 86 / \textbf{108} / 145 \\
Illinois & 489 & \scen{0.8}{3.4}{7.3} & \scen{3.8}{4.5}{5.8} & \scen{4.5}{7.9}{13.1} & 9 / \textbf{16} / 27 \\
North Carolina & 406 & \scen{0.6}{2.8}{6.1} & \scen{4.0}{4.8}{6.2} & \scen{4.6}{7.6}{12.2} & 11 / \textbf{19} / 30 \\
Nevada & 375 & \scen{0.6}{2.6}{5.6} & \scen{1.2}{1.4}{1.8} & \scen{1.8}{4.0}{7.4} & 5 / \textbf{11} / 20 \\
Wyoming & 322 & \scen{0.5}{2.2}{4.8} & \scen{3.5}{4.1}{5.3} & \scen{4.0}{6.4}{10.2} & 12 / \textbf{20} / 32 \\
\bottomrule
\end{tabular}
\end{adjustbox}
\caption{Top 15 states ranked by installed hyperscale capacity. Scenario intervals are efficiency / baseline / high-load.}
\label{tab:states_supp}
\end{table}

\begin{table}[p]
\centering
\scriptsize
\setlength{\tabcolsep}{4pt}
\begin{adjustbox}{max width=\textwidth}
\begin{tabular}{l r l l l l}
\toprule
BA & MW & Scope 1 (GL/yr) & Scope 2 (GL/yr) & Total (GL/yr) & Total (ML/MW-yr) \\
\midrule
PJM & 7,999 & \scen{12.6}{55.5}{119.6} & \scen{61.7}{74.0}{95.3} & \scen{74.2}{129.5}{215.0} & 9 / \textbf{16} / 27 \\
MISO & 2,116 & \scen{3.3}{14.7}{31.7} & \scen{14.5}{17.4}{22.4} & \scen{17.8}{32.1}{54.1} & 8 / \textbf{15} / 26 \\
SPP & 1,555 & \scen{2.4}{10.8}{23.3} & \scen{10.4}{12.5}{16.1} & \scen{12.8}{23.3}{39.3} & 8 / \textbf{15} / 25 \\
PACW & 1,220 & \scen{1.9}{8.5}{18.2} & \scen{40.0}{48.0}{61.8} & \scen{41.9}{56.5}{80.1} & 34 / \textbf{46} / 66 \\
BPA & 1,214 & \scen{1.9}{8.4}{18.2} & \scen{69.9}{83.9}{108.0} & \scen{71.8}{92.3}{126.2} & 59 / \textbf{76} / 104 \\
ERCOT & 1,193 & \scen{1.9}{8.3}{17.8} & \scen{4.8}{5.7}{7.4} & \scen{6.7}{14.0}{25.2} & 6 / \textbf{12} / 21 \\
SOCO & 699 & \scen{1.1}{4.8}{10.5} & \scen{5.6}{6.8}{8.7} & \scen{6.7}{11.6}{19.2} & 10 / \textbf{17} / 27 \\
TVA & 480 & \scen{0.8}{3.3}{7.2} & \scen{7.5}{9.0}{11.5} & \scen{8.2}{12.3}{18.7} & 17 / \textbf{26} / 39 \\
SRP & 460 & \scen{0.7}{3.2}{6.9} & \scen{3.9}{4.7}{6.1} & \scen{4.6}{7.9}{12.9} & 10 / \textbf{17} / 28 \\
DUK & 406 & \scen{0.6}{2.8}{6.1} & \scen{4.0}{4.8}{6.2} & \scen{4.6}{7.6}{12.2} & 11 / \textbf{19} / 30 \\
AZPS & 375 & \scen{0.6}{2.6}{5.6} & \scen{2.3}{2.8}{3.6} & \scen{2.9}{5.4}{9.2} & 8 / \textbf{14} / 24 \\
NEVP & 375 & \scen{0.6}{2.6}{5.6} & \scen{1.2}{1.4}{1.8} & \scen{1.8}{4.0}{7.4} & 5 / \textbf{11} / 20 \\
GCPD & 358 & \scen{0.6}{2.5}{5.4} & \scen{31.0}{37.3}{48.0} & \scen{31.6}{39.7}{53.3} & 88 / \textbf{111} / 149 \\
WACM & 322 & \scen{0.5}{2.2}{4.8} & \scen{3.5}{4.1}{5.3} & \scen{4.0}{6.4}{10.2} & 12 / \textbf{20} / 32 \\
SCEG & 310 & \scen{0.5}{2.2}{4.6} & \scen{2.5}{3.0}{3.8} & \scen{3.0}{5.1}{8.5} & 10 / \textbf{17} / 27 \\
\bottomrule
\end{tabular}
\end{adjustbox}
\caption{Top 15 balancing authorities ranked by installed hyperscale capacity. Scenario intervals are efficiency / baseline / high-load.}
\label{tab:ba_supp}
\end{table}

\end{document}